\documentclass[a4paper,fleqn]{cas-dc}

\usepackage[numbers]{natbib}
\usepackage{subfigure}

\def\tsc#1{\csdef{#1}{\textsc{\lowercase{#1}}\xspace}}
\tsc{WGM}
\tsc{QE}
\tsc{EP}
\tsc{PMS}
\tsc{BEC}
\tsc{DE}

\begin{document}
\let\WriteBookmarks\relax
\def\floatpagepagefraction{1}
\def\textpagefraction{.001}
\shortauthors{Meng Qi et~al.}

\title [mode = title]{Compact $sssc\bar{c}$ pentaquark states predicted by a quark model}

\author[1]{Qi Meng}
\cormark[1]
\ead{dz1622024@smail.nju.edu.cn}

\author[2,3,4,5]{Emiko Hiyama}

\author[3]{Kadir Utku Can}

\author[4]{Philipp Gubler}

\author[3,4]{Makoto Oka}

\author[3,4,5]{Atsushi Hosaka}

\author[1,6,7]{Hongshi Zong}

\address[1]{Department of Physics, Nanjing University, Nanjing 210093, China}
\address[2]{Department of Physics, Kyushu University, Fukuoka 819-0395, Japan}
\address[3]{Nishina Center for Accelerator-Based Science, RIKEN, Wako 351-0198, Japan}
\address[4]{Advanced Science Research Center, Japan Atomic Energy Agency, Tokai, Ibaraki 319-1195, Japan}
\address[5]{Research Center for Nuclear Physics, Osaka University, Ibaraki, Osaka 567-0047, Japan}
\address[6]{Joint Center for Particle, Nuclear Physics and Cosmology, Nanjing 210093, China}
\address[7]{State Key Laboratory of Theoretical Physics, Institute of Theoretical Physics, CAS, Beijing 100190, China}

\cortext[cor1]{Corresponding author}

\begin{abstract}

Several compact $sssc\bar c$ pentaquark resonances are predicted in a potential quark model.
The Hamiltonian is the best available one, which reproduces the masses of the low-lying charmed and strange hadrons well.
Full five-body calculations are carried out by the use of the Gaussian expansion method, and the relevant baryon-meson thresholds are taken into account explicitly. 
Employing the real scaling method, we predict four sharp resonances, 
$J^P=1/2^-$ ($E=5180$ MeV, $\Gamma=20$ MeV),
$5/2^-$ (5645 MeV, 30 MeV), 
$5/2^-$ (5670 MeV, 50 MeV), and
$1/2^+$ (5360 MeV, 80 MeV).
These are the candidates of compact pentaquark resonance states 
from the current best quark model, which should be confirmed either 
by experiments or lattice QCD calculations.

\end{abstract}


\maketitle

\section{Introduction}

Observations of candidates of multi-quark hadrons 
such as tetraquarks $X,Y,Z$~\cite{x3872,Hosaka:2016pey} and pentaquarks $P_c$~\cite{Aaij:2015tga, pc2019} gave a great impact to hadron physics community and drove  many theoretical discussions.
There have been various suggestions for their structures, here in particular for $P_c$; compact multi-quarks~\cite{Maiani:2015vwa,Lebed:2015tna,Wang:2015epa, Li:2015gta, Takeuchi:2016ejt}, hadronic molecules~\cite{Roca:2015dva,He:2015cea, Xiao:2015fia,Chen:2015loa, Liu:2019tjn}, their admixtures~\cite{Yamaguchi:2017zmn} and even baryocharmonium~\cite{Kubarovsky:2015aaa}.  
The well established $X(3872)$ and recently observed narrow pentaquark states $P_c$'s (4312, 4440, 4457) are widely expected to emerge as hadronic molecules of long range nature.
Yet compact multiquark structure with quark dynamics is an important issue to be investigated when the molecular picture can not explain high energy production processes~\cite{Esposito:2015fsa}.  
In this paper we address this question in a quark model solved by the latest advanced few-body method.  

The model we employ is the constituent quark model which accommodates important dynamics of quarks; color confinement and color magnetic spin-dependent interactions.
Hadrons are then made of minimum numbers of valence quarks.
Incorporating non-relativistic kinetic and potential energies in its Hamiltonian, the model has successfully explained many properties of low-lying conventional hadrons including their quantum numbers, masses and even interactions.  

For multi-quark states, however, the situation changes dramatically not only because of more degrees of freedom but also due to couplings to fall-apart (scattering)  channels.  
The latter occurs because multiquarks can be decomposed into more than one color singlet subsets.  
Considering these aspects two of  the present authors (E. H. and  A. H.) and collaborators studied the pentaquark systems corresponding to $\Theta^+$~\cite{hiyama2006five} and  $P_c$~\cite{hiyama2018five} in the constituent quark model 
with the scattering channels in the energy region of the observed states taken into account.  
They, however, did not find states in the experimentally observed region, but a few narrow states at significantly higher energies.  

In these studies, it was found that the coupling to the scattering states is crucially important; many states that could be found in the absence of the coupling disappear 
when scattering states are taken into account. 
It was also found that the surviving narrow states with higher energies had a spatially compact structure with little coupling to any scattering states.  
Hence the five-body analysis considering the fall-apart dynamics is essentially important.  
Yet, other features which are difficult to implement and were not considered, are those of pion dynamics; the pion exchange force and pion emission decays.  
The former is important for the formation of hadronic molecules with spatially extended structure, the latter appear as three-body decays.  

Knowing these merits and demerits of the five-body method, we propose to study the pentaquark state of $sssc\bar c$.  
Because of the flavor contents without $u,d$ quarks, the coupling to the pion can be expected to be suppressed.  
Possible meson exchange is also suppressed due to their heavier flavor contents such as $s\bar s$ or $s \bar c$.  
Moreover, thresholds of three-body open channels containing strange hadrons appear about 500 MeV above the lowest two-body ones, significantly larger than 200 MeV for the 
decays accompanying the pion. 
This work's focus on the $sss c\bar{c}$ system is furthermore promising for future comparisons between the quark model and lattice QCD calculations~\cite{Padmanath:2019wid}, since a lattice calculation would have lowered computational costs due to the absence of the light ($u$ and $d$) valance quarks. In addition, as the quark model calculation is performed at finite volume, the guidance this work provides can be helpful to understand the finite volume lattice spectrum better. A related lattice QCD study is currently underway.

This paper is organized as follows. 
After the introduction, the Hamiltonian and the employed computational method are discussed in Secs.~\ref{hami} and \ref{method}, respectively.
In Sec.~\ref{results}, we discuss our results and give a summary in Sec.~\ref{summary}.

\section{Model Hamiltonian}\label{hami}

\begin{center}
\linespread{1.3}
\begin{table}
\caption{The two parameter sets of the employed quark-quark interaction, AP1 and AL1 \cite{silvestre1996spectrum}.}\label{parameters}
\begin{tabular}{p{2.0cm}<{\centering} p{1.8cm}<{\centering} p{1.8cm}<{\centering}}
\toprule
&AP1&AL1\\
\midrule
$p$&2/3&1\\
$m_{u,d}$(GeV) & 0.277 & 0.315 \\
$m_s$(GeV) & 0.553 & 0.577 \\
$m_c$(GeV) & 1.819 & 1.836 \\
$\kappa$ & 0.4242 & 0.5069 \\
$\kappa'$ & 1.8025 & 1.8609 \\
$\lambda$(GeV$^{p+1}$) & 0.3898 & 0.1653 \\
$\Lambda$(GeV) & 1.1313 & 0.8321 \\
$B$ & 0.3263 & 0.2204 \\
$A$(GeV$^{B-1}$) & 1.5296 & 1.6553 \\
\bottomrule
\end{tabular}
\end{table}
\end{center}

\begin{center}
\linespread{1.3}
\begin{table}
\caption{The calculated masses (in MeV) of the mesons and baryons relevant for the thresholds to be considered together with the experimental values.}\label{masses1}
\begin{tabular}{p{1.1cm}<{\centering} p{1.1cm}<{\centering} p{1.3cm}<{\centering} p{1.3cm}<{\centering} p{1.3cm}<{\centering}}
\toprule
Hadron&$J^P$ & Exp. & AP1 & AL1 \\
\midrule
$\eta_c$&$0^-$& 2984 & 2984 & 3007  \\
$J/\psi$&$1^-$& 3097 & 3104 & 3103  \\
$D_s$&$0^-$& 1968 & 1955 & 1963  \\
$D_s^*$&$1^-$& 2112 & 2107 & 2102  \\
$\Omega$&$3/2^+$& 1672 & 1673 & 1675  \\
$\Omega_c$&$1/2^+$& 2695 & 2685 & 2679  \\
$\Omega_c^*$&$3/2^+$& 2766 & 2759 & 2752  \\
\bottomrule
\end{tabular}
\end{table}
\end{center}

\begin{figure*}
\centering
\includegraphics[width=1.0\textwidth]{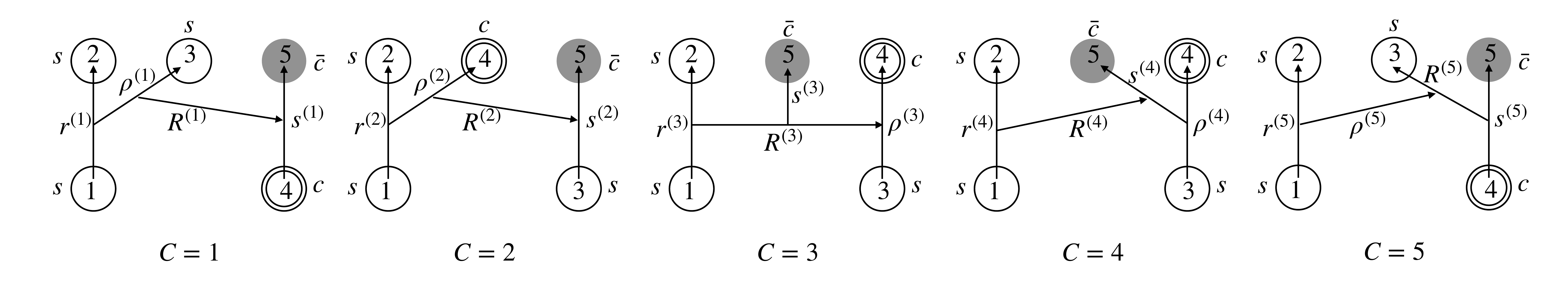}
\caption{Five sets of the Jacobi coordinate systems. The $s$ quarks, labeled as $1-3$, are to be antisymmetrized, while particles $4$ and $5$ stand for $c$ quark and $\bar{c}$ quark, respectively.
 Scatterings of $sss+c\bar{c}$ and $ssc+s\bar{c}$ are described in the coordinate bases $C=1$ and $2$, respectively.}\label{Jacobi}
\end{figure*}

The Hamiltonian of the non-relativistic quark model is given by

\begin{eqnarray}
\begin{aligned}
	H=&\sum_{i}^{5}\Big(m_i+\frac{{\boldsymbol{p}_i}^{2}}{2m_i}\Big)-T_G \\
	&-\frac{3}{16}\sum_{i<j=1}^{5}\sum_{a}^{8}\Big((\lambda_i^a \cdot \lambda_j^a) V_{ij}(\boldsymbol{r_{ij}})\Big),
\end{aligned}
\end{eqnarray}
where $m_i$ and $\boldsymbol{p}_i$ are the mass and momentum of the $i^{th}$ quark, respectively. $T_G$ is the kinetic energy of the center-of-mass motion. 
$\lambda_i^a$ are the color SU(3) Gell-mann matrices for the $i^{th}$ quark with color index $a$. We label the strange quarks, $s$ as $i=1,2,3$, the charm quark, $c$ as $i=4$, and the anticharm quark $\bar{c}$ as $i=5$.

We use the quark-quark interaction potential proposed by Semay and Silvestre-Brac \cite{silvestre1996spectrum, semay1994diquonia}, given by

\begin{eqnarray}
\begin{aligned}
	V_{ij}(\boldsymbol{r})=&-\frac{\kappa}{r}+\lambda r^p-\Lambda \\
	&+\frac{2\pi\kappa'}{3m_i m_j}\frac{\mathrm{exp}(-r^2/r_0^2)}{\pi^{3/2}r_0^3}\boldsymbol{\sigma}_i\cdot\boldsymbol{\sigma}_j,
\end{aligned}
\end{eqnarray}
with
\begin{eqnarray}
r_0(m_i,m_j)=A(\frac{2m_i m_j}{m_i+m_j})^{-B}.
\end{eqnarray}

This potential consists of the color Coulomb potential, the linear confining part, a  (color-electric) constant term and the color-magnetic spin-spin interaction term. The last term comes from a magnetic gluon exchange, where the $\delta$ function in the Breit-Fermi interaction is modified by a cutoff parameter $r_0$. Note that $r_0$ depends on the reduced quark masses. The two sets of parameter choices appearing in this work, AP1 and AL1, are listed in Table~\ref{parameters}.

The present Hamiltonian is tested by computing the static properties of low-lying baryons and mesons. 
The calculated masses are given in Table~\ref{masses1} for the AP1 and AL1 parameters together with the corresponding experimental values. 
In one earlier work, this Hamiltonian was used in a pentaquark system ($qqqc\bar{c}$) calculation [15]. We choose AP1 in our present calculation since it reproduces the relevant thresholds better. In addition, we have tested the AL1 in our five-body calculation also and have confirmed that the results are not qualitatively modified by this alternative choice.

\section{Method}\label{method}

In this section, we briefly discuss our method of numerically solving the five-body Schr\"{o}dinger equation. 
We describe the five-body wave function with five types of Jacobi coordinates shown in Fig.~\ref{Jacobi}.
$C=1$ and $2$ are configurations in which two color-singlet clusters may fall apart along the inter-cluster coordinates $\boldsymbol{R}^{(c)}(C=1,2)$ . Namely, for $C=1$, the color wave function is chosen as the product of color-singlet $sss$ plus $c\bar{c}$, which correspond to $\eta_c\Omega$ and $J/\psi\Omega$ configurations. For $C=2$, the color wave function is chosen as the product of color-singlet $ssc$ plus $s\bar{c}$, which correspond to $D_s\Omega_c$, $D_s^*\Omega_c$, $D_s\Omega_c^*$, and $D_s^*\Omega_c^*$ configurations. In contrast, the other three configurations, $C=3-5$, do not describe color-singlet subsystems, and represent the five quarks as always connected by a confining interaction. In this sense, we call $C=3-5$ as the "connected" (confining) configurations.

The five-body Schr\"{o}dinger equation for the total angular momentum $J$ and its z-component $M$ is given by
\begin{eqnarray}
(H-E)\Psi_{JM}=0.
\end{eqnarray}

We solve it by using the Gaussian Expansion Method (GEM) \cite{GEM1988,GEM2003}, which was successfully applied to various types of three-body and four-body systems \cite{hiyama2000lambda, hiyama2004four, hiyama2009structure, hiyama2015resonant, brink1998tetraquarks}.
The total wave function $\Psi_{JM}$ is written as a sum of components, each described in terms of one of the Jacobi coordinate bases, 

\begin{eqnarray}
\begin{aligned}
	\Psi_{JM}=&\sum_C \mathcal{A}_{123}\xi_1^{(C)} \sum_{\gamma} B_{\gamma}^{(C)}  \\
	&\times \bigg[\chi^{(C)}_{S} \Phi_{L}^{(C)}(r^{(C)},\rho^{(C)},R^{(C)},s^{(C)}) \bigg]_{JM},
\end{aligned}
\label{total wave function}
\end{eqnarray}
where $C$ specifies the set of Jacobi coordinates. $\xi_1^{(C)}$, $\chi^{(C)}_{S}$, and $\Phi_{L}^{(C)}$ represent the color-singlet wave functions, spin wave functions for total spin $S$ and spatial wave functions for total orbital angular momentum $L$, respectively. $\mathcal{A}_{123}$ denotes the anti-symmetrization operator for the three $s$ quarks (1,2,3).
 
The color-singlet total wave functions, $\xi_1^{(C)}$, for $C=1-5$ are chosen as
\begin{eqnarray}
&&\xi_1^{(1)}=[(123)_1(45)_1]_1,	\nonumber\\
&&\xi_1^{(2)}=[(124)_1(35)_1]_1,	\nonumber\\
&&\xi_1^{(3)}=[[(12)_{\bar{3}}(34)_{\bar{3}}]_3 5]_1	,	\nonumber\\
&&\xi_1^{(4)}=[(12)_{\bar{3}}[(34)_{\bar{3}}5]_3]_1,		\nonumber\\
&&\xi_1^{(5)}=[(12)_{\bar{3}}[(45)_1 3]_3]_1.
\end{eqnarray}

The spin wave functions for the total spin $S$ are given by 
\begin{eqnarray}
&&\chi^{(1)}_{S}=[[(12)_s 3]_{\sigma}(45)_{\bar{s}}]_{S}	,	\nonumber\\
&&\chi^{(2)}_{S}=[[(12)_s 4]_{\sigma}(35)_{\bar{s}}]_{S}	,	\nonumber\\
&&\chi^{(3)}_{S}=[[(12)_s(34)_{\bar{s}}]_{\sigma}5]_S,		\nonumber\\
&&\chi^{(4)}_{S}=[(12)_s[(34)_{\bar{s}}5]_{\sigma}]_S,	\nonumber\\
&&\chi^{(5)}_{S}=[(12)_s[(45)_{\bar{s}}3]_{\sigma}]_S,
\end{eqnarray}
where $s$, $\bar{s}$, and $\sigma$ represent the spins of the subsystem designated in each definition.

The spatial wave function is expanded by Gaussian basis functions as
\begin{eqnarray}
\begin{aligned}
\Phi_{L}^{(C)}(r^{(C)},&\rho^{(C)},R^{(C)},s^{(C)})= \\
& \bigl[ \phi(n_1,l_1,m_1,\boldsymbol{r}^{(C)}) \times \phi(n_2,l_2,m_2,\boldsymbol{\rho}^{(C)}) \\
\times &\phi(n_3,l_3,m_3,\boldsymbol{R}^{(C)})
\times \phi(n_4,l_4,m_4,\boldsymbol{s}^{(C)}) \bigr]_L,
\end{aligned}
\end{eqnarray}	
where $\phi(n,l,m,\boldsymbol{r})$ is defined as 
\begin{eqnarray}
	\phi(n,l,m,\boldsymbol{r})=N_{nl}r^l e^{-(r/r_n)^2}Y_{lm}(\hat{\boldsymbol{r}})
\end{eqnarray}
with the Gaussian ranges taken in geometric progression,
\begin{eqnarray}
	r_n=r_1 a^{n-1} \qquad (n=1-n_{max}).
\end{eqnarray}

The $\gamma$ index in the total wave function $\Psi_{JM}$ given in Eq.~\ref{total wave function} is defined as
\begin{eqnarray}
\gamma \equiv \{s,\bar{s},\sigma,S,n_1,n_2,n_3,n_4,l_1,l_2,l_3,l_4,L \}.
\end{eqnarray}
For completeness, we note that the orbital angular momenta are combined in the order of 
$(((l_1,l_2) ,l_3),l_4)L$, where the intermediate quantum numbers are suppressed.
Within the present calculation settings, they are determined uniquely, so that we can omit them.

The dimensions of the basis of Gaussian wave functions, $n_{1max}$, $n_{2max}$, $n_{4max}$ for the $C=1-5$ channels are 6. $n_{3max}$ for $C=1$ and $2$ equals 10, and for $C=3-5$ are set to 6. 

In the present calculation, we investigate both positive and negative parity states.
For the negative parity states, we take the total orbital angular momentum as $L=0$ and the total spin-parity as $J^P=1/2^-$, $3/2^-$, and $5/2^-$. 
The orbital angular momenta of $\boldsymbol{r}$, $\boldsymbol{\rho}$, $\boldsymbol{R}$, and $\boldsymbol{s}$ for $C=1$ and $2$ are chosen as $(l_1,l_2,l_3,l_4)=(0,0,0,0)$, $(1,0,0,1)$, and $(0,1,0,1)$. 
For $C=3-5$, we set all the orbital angular momenta to 0.
For the positive parity states, the total orbital angular momentum is taken to $L=1$ and the total spin-parity to $J^P=1/2^+$, $3/2^+$, and $5/2^+$. 
The orbital angular momenta of $\boldsymbol{r}$, $\boldsymbol{\rho}$, $\boldsymbol{R}$, and $\boldsymbol{s}$ for $C=1$ and $2$ are 
chosen as $(l_1,l_2,l_3,l_4)=(0,0,1,0)$, $(0,0,0,1)$, $(1,0,0,0)$, and $(0,1,0,0)$, and for $C=3-5$ are chosen as $(l_1,l_2,l_3,l_4)=(0,0,1,0)$.

In diagonalizing the five-body Hamiltonian, we use about 40,000 basis functions for $J^P=1/2^-$, $3/2^-$, $1/2^+$, $3/2^+$, and 15,000 basis functions for $J^P=5/2^-$ and $5/2^+$.

It should be noted here that all the obtained eigenvalues are discrete. 
Namely, as the system is computed in a finite volume, 
even the continuum states corresponding to the baryon-meson scattering solutions come out as discrete states. 
In earlier works \cite{hiyama2006five,hiyama2018five} , the real-scaling (stabilization) method \cite{1981stablization} was adopted to distinguish genuine resonances from the discretized scattering states.
In the present case, we scale the basis functions along $\boldsymbol{R}^{(1)}$ and $\boldsymbol{R}^{(2)}$ by multiplying all the range parameters simultaneously with a factor $\alpha$ as $R_N \rightarrow \alpha R_N $. 
Then, any continuum state will fall off towards its threshold, while a compact resonant state should not be affected by the boundary at a large distance.

\section{Results}\label{results}

\begin{figure*}
\centering
\includegraphics[width=0.95
\textwidth]{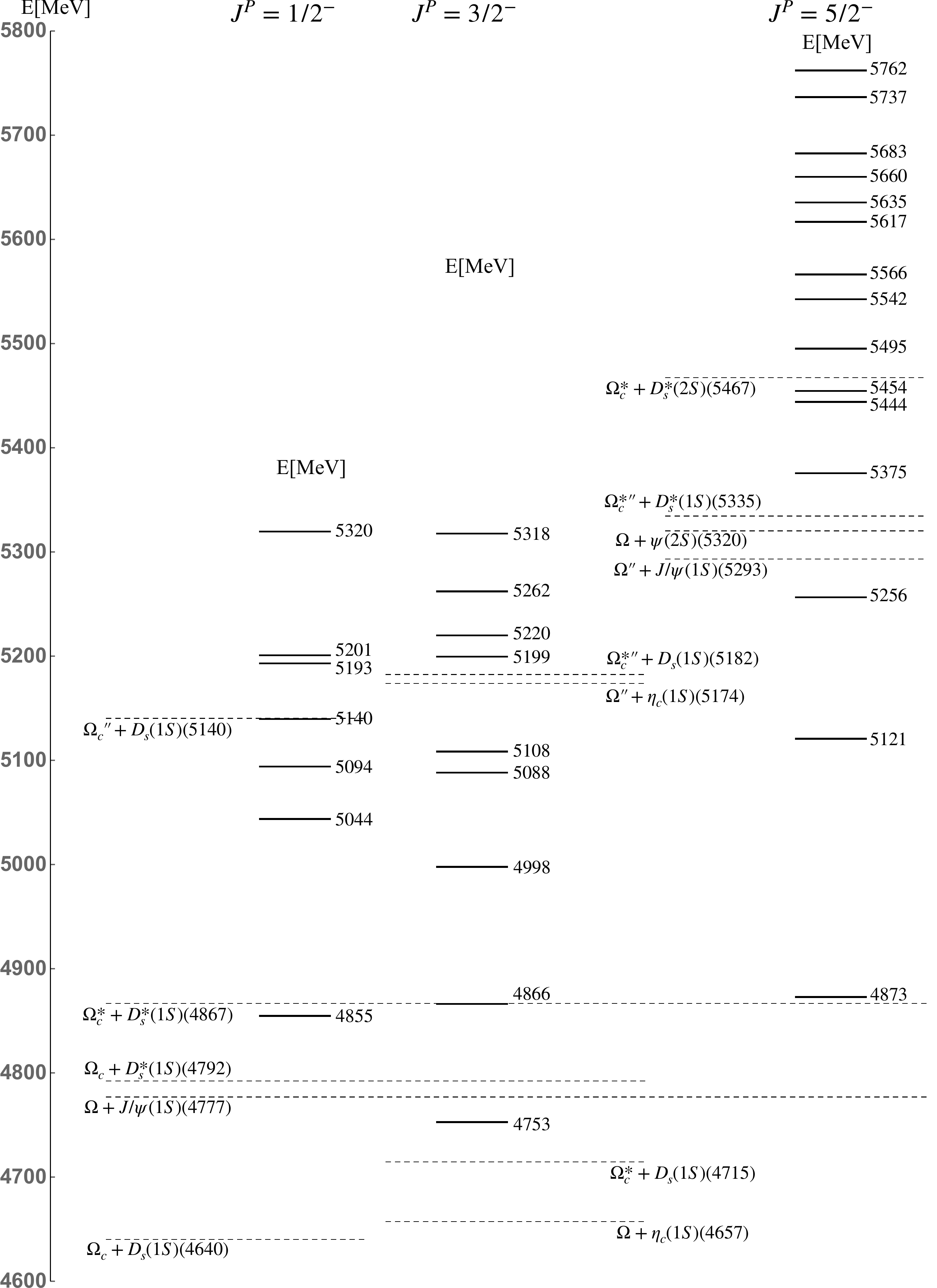}
\caption{The calculated energy spectra for quantum numbers $J^P=1/2^-,3/2^-,5/2^-$, including only the connected configurations $C=3-5$ 
are shown in units of MeV. The dashed lines are thresholds, drawn according to the theoretical numbers given in Tables.~\ref{masses1} and \ref{masses2}.}\label{connected}
\end{figure*}

\begin{figure*}
\centering
\includegraphics[width=0.95
\textwidth]{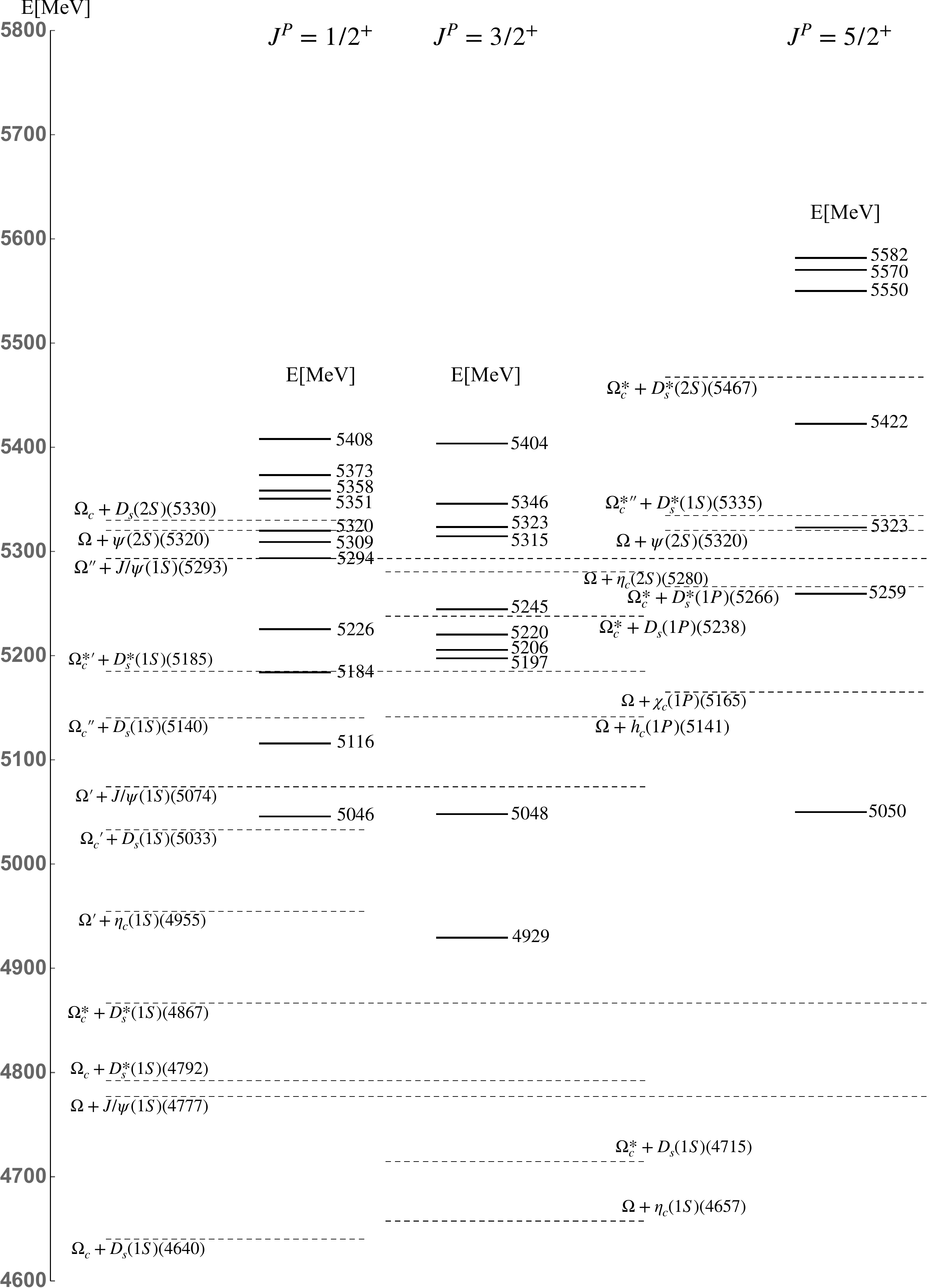}
\caption{Same as in Fig.~\ref{connected}, but for $J^P=1/2^+,3/2^+,5/2^+$.}\label{connectedplus}
\end{figure*}

As a first step, we calculated the spectra without the contributions from scattering states. 
In these calculations, only the connected Jacobi coordinate bases in Fig.~\ref{Jacobi}, $C=3-5$ are included. 
Solving the Schr\"{o}dinger equation for $J^P=1/2^-$, $3/2^-$, $5/2^-$, $1/2^+$, $3/2^+$, and $5/2^+$, we obtain the spectra shown in Figs.~\ref{connected} and \ref{connectedplus}. 
It can be seen in these figures that all eigenvalues are obtained above the lowest meson-baryon thresholds. 
Because these calculations include only the connected channels $C=3-5$ without contributions from the scattering channels $C=1$ and $2$, all states are stable against fall-apart decay. We call these states compactified states in the following because they are forces to be so.

In Figs.~\ref{connected} and \ref{connectedplus}, the dashed lines indicate the relevant meson-baryon thresholds, which couple to the shown compactified states. 
The calculated masses of the excited mesons and baryons corresponding to the thresholds, are given in Table~\ref{masses2} for the AP1 parameters. 
Here, 
$\Omega'$ stand for the first excited states with negative parity, while $\Omega''$ represent the positive-parity excited states of $\Omega$ baryons. 
As we neglect the spin-orbit interaction in the current Hamiltonian, the p-wave mesons, $\chi_c(1P) (J^{P}=0^+, 1^+, 2^+)$ and  $D_s^*(1P) (0^+, 1^+,\\ 2^+)$, are degenerate. The same happens for $\Omega'(1/2^-,3/2^-)$, $\Omega_c'(1/2^-,3/2^-)$, and $\Omega_c^{*}{'}(1/2^-,3/2^-)$. 
Note that the lowest and second thresholds in our calculation [$\Omega_c+D_s(1S)$ and $\Omega+\eta_c(1S)$] are reversed from the experimental data, which give $\Omega_c+D_s(1S)  (4663)$ and $\Omega+\eta_c(1S) (4656)$.

\begin{center}
\linespread{1.2}
\begin{table}
\caption{The calculated masses (in MeV) of the excited mesons and baryons relevant to the thresholds to be considered.}\label{masses2}
\begin{tabular}{p{1.5cm}<{\centering} p{1.8cm}<{\centering} p{1.5cm}<{\centering}}
\toprule
Hadron&$J^P$ & AP1 \\
\midrule
 $h_c(1P)$  & $1^+$ & 3468    \\
 $\eta_c(2S)$  & $0^-$ & 3607    \\
 $\chi_c(1P)$     & $0^+$,$1^+$,$2^+$ & 3492    \\
 $\psi(2S)$    & $1^-$ & 3647    \\
 $D_s(1P)$     & $1^+$ & 2479    \\
 $D_s(2S)$     & $0^-$ & 2648    \\
 $D_s^{*}(1P)$ & $0^+$,$1^+$,$2^+$ & 2507    \\
 $D_s^{*}(2S)$ & $1^-$ & 2708    \\
 $\Omega'$     & $1/2^-$,$3/2^-$ & 1971    \\
 $\Omega''$    & $3/2^+$ & 2190    \\
 $\Omega_c'$   & $1/2^-$,$3/2^-$ & 3078    \\
 $\Omega_c''$     & $1/2^+$ & 3185    \\
 $\Omega_c^{*}{'}$  & $1/2^-$,$3/2^-$ & 3078    \\
 $\Omega_c^{*}{''}$ & $3/2^+$ & 3227    \\ 

\bottomrule
\end{tabular}
\end{table}
\end{center}

For spin-parity $J^P=1/2^-$, the lowest energy state appears at 4855 MeV, which is above the hidden charm threshold $\Omega+J/\psi$ and open charm thresholds $\Omega_c+D_s$ and $\Omega_c+D_s^*$, 
but below the open charm threshold $\Omega_c^*+D_s^*$. The second state is located at 5044 MeV. 

For spin-parity $J^P=3/2^-$, the lowest eigenvalue is obtained at 4753 MeV. The second state appears at 4866 MeV which is slightly higher than the lowest state of $J^P=1/2^-$, while the third one shows up at 4998 MeV. 

For spin-parity $J^P=5/2^-$, the lowest state is found at 4873 MeV which is higher than all the open charm thresholds and hidden charm thresholds formed by ground states. 

The lowest levels of the $J^P=1/2^+$ and $5/2^+$ channels are located at 5046 MeV and 5050 MeV, respectively. For $J^P=3/2^+$, the lowest eigenvalue appears at 4929 MeV.
Compared to the negative parity states, the lowest energy levels of positive parity are located above all the thresholds which consist of only ground state baryons and ground state mesons.

Next, we consider the contributions from scattering states by including the scattering channels $C=1$ and $2$ in our calculation. According to Ref.~\cite{hiyama2018five}, 
the coupling of the scattering states may cause some of the compactified states to melt into the continuum spectrum. 

To investigate the nature of each compactified state, we include the scattering state one by one in the real scaling method calculation. Namely, we scale the range 
parameter $\boldsymbol{R}_N$ of the Gaussian bases as $\boldsymbol{R}_N \rightarrow \alpha \boldsymbol{R}_N$ for the scattering channel in the Jacobi coordinates of $C=1$ or $2$. 
The eigenvalues corresponding to scattering states will fall down towards the respective thresholds with the increasing $\alpha$ values. At the same time, 
the resonant states will stay at their energy independently from the scaling factor $\alpha$. With this procedure, we can determine the dominant meson-baryon component for each compactified state.
For more details and examples, see Ref.~\cite{hiyama2018five}.

\linespread{1.0}
\begin{table}
\centering
\caption{Dominant Baryon-Meson components for the various compactified states in Fig.~\ref{connected} for the $J^P=1/2^-,3/2^-, 5/2^-$ channels.}\label{dominantn}
\begin{tabular}{p{1.5cm}<{\centering} p{1.0cm}<{\centering} p{4.6cm}<{\centering}}
\toprule
$J^P=1/2^-$& energy (MeV) & configuration\\
\midrule
 & 4855 & $\Omega+J/\psi(1S)$, $\Omega_c+D_s(1S)$ \\
 & 5044 & $\Omega_c+D_s(1S)$, $\Omega_c+D_s^{*}(1S)$ \\
 & 5094 & $\Omega+J/\psi(1S)$, $\Omega_c^*+D_s^{*}(1S)$ \\
 & 5140 & $\Omega+J/\psi(1S)$ \\
 & 5193 & $\Omega_c''+D_s(1S)$, $\Omega_c+D_s^{*}(1S)$ \\
 & 5201 & - \\
 & 5320 & - \\
\midrule
$J^P=3/2^-$& energy (MeV) & configuration\\
\midrule
 & 4753 & $\Omega+\eta_c(1S)$, $\Omega_c^{*}+D_s(1S)$  \\
 & 4866 & $\Omega+J/\psi(1S)$, $\Omega_c^{*}+D_s(1S)$ \\
 & 4998 & $\Omega+\eta_c(1S)$, $\Omega_c^{*}+D_s(1S)$  \\
 & 5088 & $\Omega_c^{*}+D_s^{*}(1S)$  \\
 & 5108 & $\Omega+J/\psi(1S)$, $\Omega_c+D_s^{*}(1S)$ \\
 & 5199 & $\Omega''+\eta_c(1S)$, $\Omega_c^{*}{''}+D_s(1S)$, $\Omega_c^{*}+D_s^{*}(1S)$  \\
 & 5220 & $\Omega+J/\psi(1S)$  \\
 & 5262 & $\Omega''+\eta_c(1S)$  \\
 & 5318 &  -  \\
\midrule
$J^P=5/2^-$& energy (MeV) & configuration\\
\midrule
 & 4873 & $\Omega+J/\psi(1S)$ \\
 & 5121 & $\Omega+J/\psi(1S)$ \\
 & 5256 & $\Omega_c^*+D_s^*(1S)$ \\
 & 5375 & $\Omega''+J/\psi(1S)$, $\Omega_c^{*}{''}+D_s^*(1S)$ \\
 & 5444 & $\Omega''+J/\psi(1S)$ \\
 & 5454 & $\Omega''+J/\psi(1S)$, $\Omega_c^{*}{''}+D_s^*(1S)$ \\
 & 5495 & $\Omega+\psi(2S)$, $\Omega_c^*+D_s^*(2S)$ \\
 & 5542 & $\Omega_c^*+D_s^*(2S)$ \\
 & 5566 & $\Omega+\psi(2S)$, $\Omega_c^*+D_s^*(2S)$ \\
 & 5617 & $\Omega+\psi(2S)$ \\
 & 5635 & $\Omega+\psi(2S)$, $\Omega_c^*+D_s^*(2S)$ \\
 & 5660 & - \\
 & 5683 & $\Omega+\psi(2S)$ \\
 & 5737 & $\Omega+\psi(2S)$, $\Omega_c^*+D_s^*(2S)$ \\
 & 5762 & - \\
\bottomrule
\end{tabular}
\end{table}

\begin{table}
\centering
\caption{Dominant Baryon-Meson components for the various compactified states in Fig.~\ref{connectedplus} for the $J^P=1/2^+,3/2^+, 5/2^+$ channels. }\label{dominantp}
\begin{tabular}{p{1.5cm}<{\centering} p{1.0cm}<{\centering} p{4.6cm}<{\centering}}
\toprule
$J^P=1/2^+$& energy (MeV) & configuration\\
\midrule
 & 5046 & $\Omega+J/\psi(1S)$, $\Omega_c+D_s(1S)$ \\
 & 5116 & $\Omega{'}+\eta_c(1S)$, $\Omega_c{'}+D_s(1S)$ \\
 & 5184 & $\Omega_c{''}+D_s(1S)$ \\
 & 5226 & $\Omega{'}+J/\psi(1S)$ \\
 & 5294 & $\Omega_c^*+D_s^*(1S)$, $\Omega_c+D_s^*(1S)$ \\
 & 5309 & $\Omega{''}+J/\psi(1S)$, $\Omega_c+D_s^*(1S)$ \\
 & 5320 & $\Omega{''}+J/\psi(1S)$, $\Omega_c+D_s^*(1S)$ \\
 & 5351 & $\Omega+\psi(2S)$, $\Omega_c^*{'}+D_s^*(1S)$ \\
 & 5358 & $\Omega{'}+\eta_c(1S)$, $\Omega_c+D_s(2S)$ \\
 & 5373 & $\Omega_c+D_s(2S)$ \\
 & 5408 & - \\
\midrule
$J^P=3/2^+$& energy (MeV) & configuration\\
\midrule
 & 4929 & $\Omega+\eta_c(1S)$ \\
 & 5048 & $\Omega+J/\psi(1S)$ \\
 & 5197 & $\Omega+h_{c}(1P)$ \\
 & 5206 & $\Omega+h_{c}(1P)$, $\Omega_c^*{'}+D_s^*(1S)$ \\
 & 5220 & $\Omega{'}+J/\psi(1S)$, $\Omega_c+D_s^*(1S)$, $\Omega_c^*+D_s(1S)$ \\
 & 5245 & $\Omega_c^*+D_s(1P)$, $\Omega_c^*{'}+D_s^*(1S)$ \\
 & 5315 & $\Omega_c^{*}+D_s^*(1S)$ \\
 & 5323 & $\Omega{''}+J/\psi(1S)$ \\
 & 5346 & $\Omega+\eta_c(2S)$ \\
 & 5404 & $\Omega+\eta_c(2S)$, $\Omega{''}+J/\psi(1S)$ \\
\midrule
$J^P=5/2^+$& energy (MeV) & configuration\\
\midrule
 & 5050 & $\Omega+J/\psi(1S)$ \\
 & 5259 & $\Omega+\chi_c(1P)$, $\Omega_c^*+D_s^*(1S)$ \\
 & 5323 & $\Omega{''}+J/\psi(1S)$ \\
 & 5422 & $\Omega+\psi(2S)$, $\Omega_c^*+D_s^*(1P)$ \\
 & 5550 & $\Omega_c^*{''}+D_s^*(1S)$ \\
 & 5570 & $\Omega+\psi(2S)$, $\Omega_c^*+D_s^*(2S)$ \\
 & 5582 & - \\
\bottomrule
\end{tabular}
\end{table}

Following the above procedure, we now study the coupling of each compactified state to specific scattering states. 
The results are summarized in Table~\ref{dominantn} for negative parity states and in Table~\ref{dominantp} for positive parity states. 
They show that most of the compactified states have significant coupling to some scattering states, and they do not survive as resonances. For instance, one sees that the lowest negative parity state, 4753 MeV ($3/2^-$)
is mainly an $\Omega+\eta_c$ scattering state, while 4855 MeV ($1/2^-$), 4866 MeV ($3/2^-$), and 4873 MeV ($5/2^-$), have a dominant overlap with $\Omega + J/\psi$ scattering state. 
The next group of excited states are similarly assigned to $\Omega_c+D_s$, $\Omega_c^*+D_s$, and $\Omega_c^*+D_s^*$ states.

After removing the scattering states, compact resonances remain at 5201 MeV and 5320 MeV for $J^P=1/2^-$, at 5318 MeV for $J^P=3/2^-$, at 5660 MeV and 5762 MeV for $J^P=5/2^-$, at 5408 MeV for $J^P=1/2^+$, 
and at 5582 MeV for $J^P=5/2^+$. 
For $J^P=3/2^+$, all the compactified states in the low energy region have dominant scattering configurations.

\begin{figure*}
\centering
\subfigure[$C=1$ and $2$]{
\begin{minipage}[b]{1.0\textwidth}
\includegraphics[width=1\textwidth]{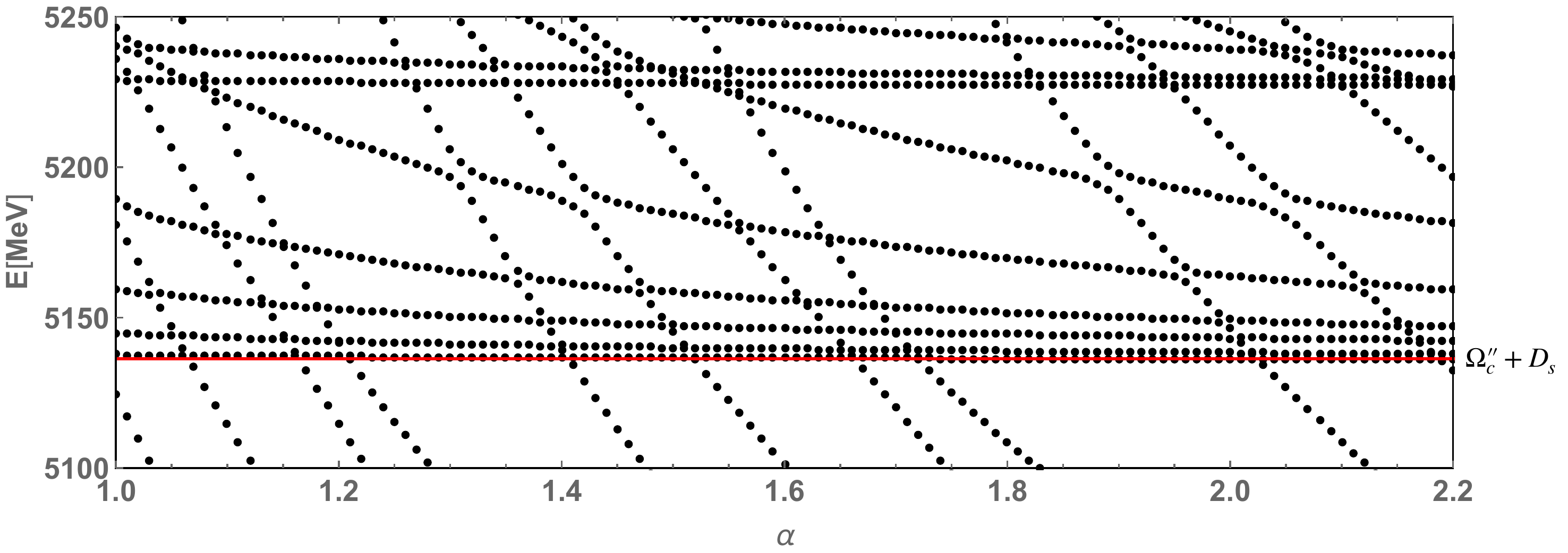}
\end{minipage}
}
\subfigure[$C=1-5$]{
\begin{minipage}[b]{1.0\textwidth}
\includegraphics[width=1\textwidth]{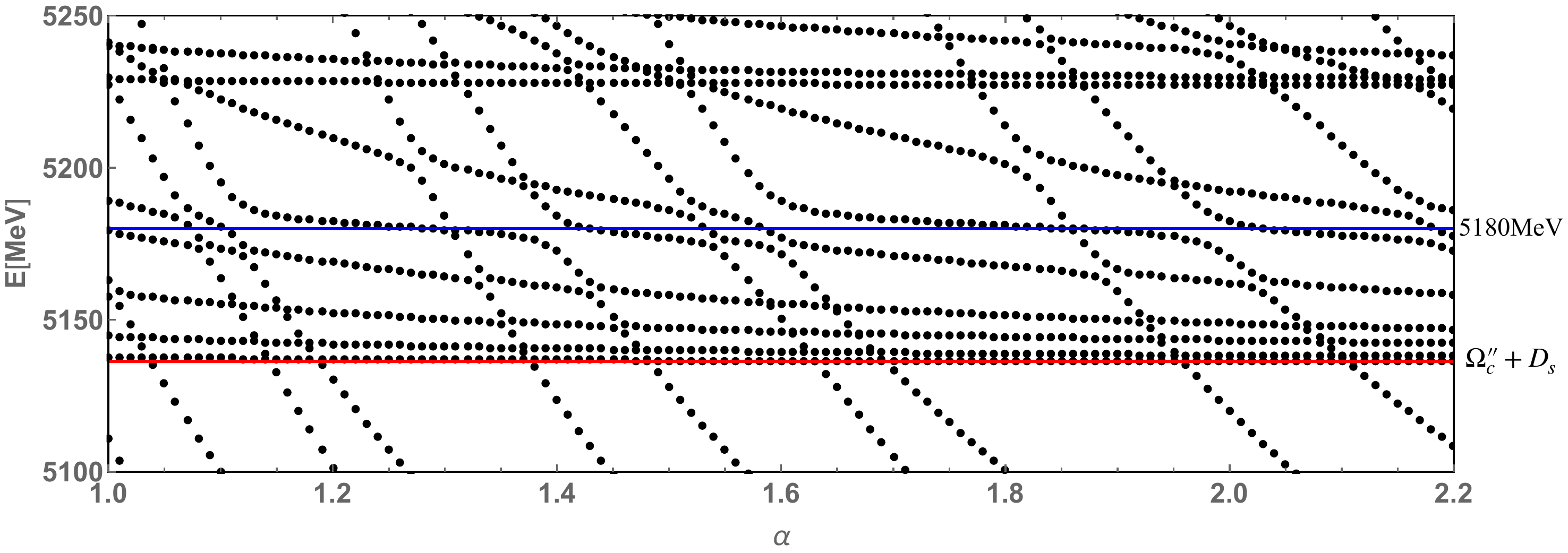}
\end{minipage}
}
\caption{The stabilization plots of the eigenenergies $E$ for $J^P=1/2^-$ with the respect to the scaling factor $\alpha$ in two cases: (a) including only scattering configurations $C=1$ and $2$; (b) including full configurations $C=1-5$. The Gaussian ranges $R_N$ for the coordinates $\boldsymbol{R}_1$ and $\boldsymbol{R}_2$ of $C=1$ and $2$ configurations are scaled as $R_N \rightarrow \alpha R_N$ with $\alpha=1.0-2.2$. The red line is threshold. The blue line is the location of 5180 MeV.}\label{12kfull}
\end{figure*}

Let us investigate one of the compact resonances in more detail. 
In Fig.~\ref{12kfull}, we show the stabilization plots using the real scaling method for the $J^P=1/2^-$ channel. Fig.~\ref{12kfull}(a) shows the results when only the scattering configurations $C=1$ and $2$ are included. 
Fig.~\ref{12kfull}(b) shows the results when all configurations $C=1-5$ are incorporated in the calculation. As one can see, around 5201 MeV, there is a clear difference between only scattering configurations and full configurations. 
By including the connected configurations $C=3-5$, a resonance structure appears at around 5180 MeV, for which the compactified state at 5201 MeV 
can be considered as a seed. With such stabilization plots, we can estimate the width of resonance states \cite{1981stablization}. The width of the one around 5180 MeV is estimated to be 20 MeV.

We note that there are two more possible configurations, namely $[[(12)_{\bar3}5]_3 (34)_{\bar3}]_1$ and 
$[(12)_{\bar3}[(35)_{1}4]_3]_1$, for the configuration sets of $C=4$ and $C=5$, respectively. Our numerical tests indicate that these additional configurations indeed belong to configuration sets $C=4$ and $C=5$, and their respective energies lay higher than that of $C=4$ and $C=5$ given in Eq.6. Therefore, these additional configurations are unlikely to be helpful in generating the lowest energy level of this system. In addition, the resonance energies from a solution incorporating all the possible configurations typically deviate by at most a few $10$ MeV from the energies obtained from a solution with the channels employing only $C=3,4,5$.

\begin{center}
\linespread{1.0}
\begin{table}
\caption{ Resonance structures, their "seed" compactified states and the estimated decay widths.}\label{resonance}
\begin{tabular}{p{1.0cm}<{\centering} p{1.8cm}<{\centering} p{1.8cm}<{\centering} p{1.8cm}<{\centering}}
\toprule
 $J^P$ & energy (MeV) & width (MeV) & "seed"(MeV) \\
\midrule
$1/2^-$ & 5180 & 20 & 5201\\
        & 5290 & $>$100 & 5320\\
$3/2^-$ & 5300 & $>$100 & 5318\\
$5/2^-$ & 5645 & 30 & 5660\\
	    & 5670 & 50 & 5762\\
$1/2^+$ & 5360 & 80 & 5408 \\
$5/2^+$ & 5570 & $>$100 & 5582 \\
\bottomrule
\end{tabular}
\end{table}
\end{center}

With the same method, we studied all other compact resonance structures with different spin-parity quantum numbers. The results are summarized in Table~\ref{resonance}.

\begin{figure}
\begin{center}
\includegraphics[width=8.0cm]{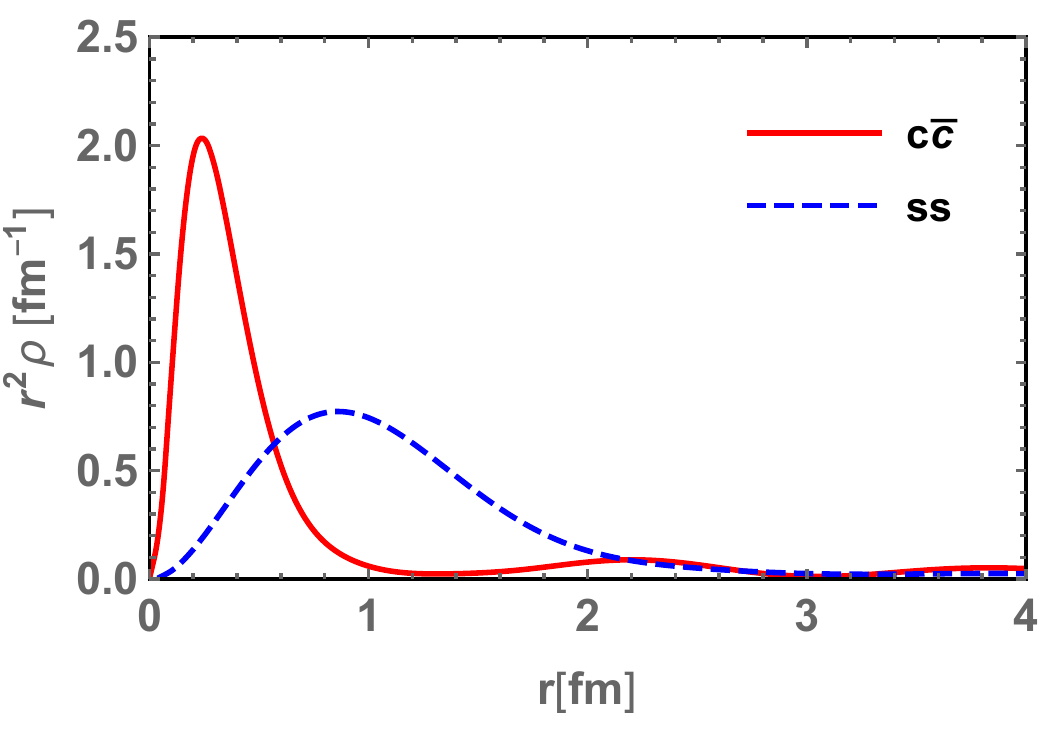}
\caption{Density distributions ${r_1}^2 \rho_{ss}(r_1)$ and ${s_1}^2 \rho_{c\bar{c}}(s_1)$ as functions of the distance $r=r_1=s_1$}\label{r2rho}
\end{center}
\end{figure}

To obtain more information 
about the spatial structures of the lowest $J^P=1/2^-$ compact resonance, 
we calculated the two-body correlation functions of $ss$ and $c\bar{c}$ for this state. The correlation functions are defined as 
\begin{eqnarray}
&&\rho_{ss}(r_1)=\int|\Psi_{JM}|^{2}d\boldsymbol{s}_1d\boldsymbol{R}_1d\boldsymbol{\rho}_1d\boldsymbol{\hat{r}}_1		\nonumber\\
&&\rho_{c\bar{c}}(s_1)=\int|\Psi_{JM}|^{2}d\boldsymbol{r}_1d\boldsymbol{R}_1d\boldsymbol{\rho}_1d\boldsymbol{\hat{s}}_1	
\end{eqnarray}
where $\boldsymbol{r}_1$ and $\boldsymbol{s}_1$ are the relative distances between $ss$ and $c\bar{c}$. $d\boldsymbol{\hat{r}}_1$ and $d\boldsymbol{\hat{s}}_1$ denote the integral of angular parts of $\boldsymbol{r}_1$ and $\boldsymbol{s}_1$, respectively.
The integral is performed at $E=5180$MeV and $\alpha=1.28$. Fig.~\ref{r2rho} shows the density distributions of ${r_1}^2 \rho_{ss}(r_1)$ and ${s_1}^2 \rho_{c\bar{c}}(r_1)$ as 
functions of the distance $r=\boldsymbol{r_1}=\boldsymbol{s_1}$. The peak position of $c\bar{c}$ is found at about 0.25 fm, which is more compact than charmonia $J/\psi$ or $\eta_c$. 
The corresponding $ss$ peak lies at about 0.85fm which is more extended than the $\Omega$ baryon. 

\section{Summary}\label{summary}

In this paper, we have studied $sssc\bar c$ pentaquarks of 
$J^P=1/2^-$, $3/2^-$, $5/2^-$, $1/2^+$, $3/2^+$, and $5/2^+$.
The potential quark model is used to analyze the spectrum and resonance energies are obtained from the most precise five-body calculation available to date.

The key findings of our calculation can be summarized as follows. 

(1) Our Hamiltonian is taken from Semay and Silvestre-Brac (SSB) \cite{silvestre1996spectrum, semay1994diquonia}.
As is shown in Tables~\ref{masses1} and \ref{masses2}, the SSB model reproduces the hadron masses within 15 MeV, which are relevant for the open channel thresholds of the current pentaquark systems. This is very important to guarantee the correctness of the quark dynamics in the strange and charm sectors, and also to compare our results with the real (observed) spectrum that is influenced strongly by the open channel thresholds. In this sense this is the best available potential for the present calculation.

(2) In order to estimate the systematic uncertainties of the model, we have compared two sets of parameter choices of the SSB potential, AP1 and AL1. The main results shown above are for the AP1 potential, which fits the observed data better than AL1. However we have found that the pentaquark resonances for AL1 come out at similar energies as AP1.
For instance, we find a sharp $1/2^-$ resonance at 5220 MeV with a width of 25 MeV (AL1), compared to  5180 MeV with a width of 20 MeV (AP1).

(3) For the given Hamiltonian, we have solved the $sssc\bar{c}$ states as precisely as possible.
The Gaussian expansion method (based on the variational principle) is employed for the full five-body system.
In our five-body calculation, we include all relevant meson-baryon scattering channels (Cf. $C=1,2$ of Fig.~\ref{Jacobi}) such as $\Omega + \eta_c$, $\Omega + J/\psi $, $\Omega_c^{*} + D_s^{*}$, and so on. As a result, we need more than 40,000 basis functions for this system.
To distinguish two-body scattering states and compact resonant states, we have employed the real scaling method (stabilization method) which has been successfully applied in previous studies \cite{hiyama2006five,hiyama2018five}.

(4) There, however, is a caveat in the current quark model. 
As neither $q\bar q$ creations nor explicit mesons are introduced, the model Hamiltonian cannot describe meson exchange interactions between hadrons. 
It was pointed out that the $P_c$ ($uudc\bar c$) pentaquarks observed by LHCb can be realized as molecular-type $\Sigma_c + \bar D$ and $\Sigma_c + \bar{D}^{*}$  resonances due to the attractive pion (meson) exchange potential.
Such resonance states may not appear in the present calculation because once the two color-singlet hadrons are separated in $C=1$, or 2 in Fig.~\ref{Jacobi}, there is no interaction between them.

Within these conditions, we predict four sharp resonances:
$J^P=1/2^-$ ($E=5180$ MeV, $\Gamma=20$ MeV),
$5/2^-$ (5645 MeV, 30 MeV), 
$5/2^-$ (5670 MeV, 50 MeV), and
$1/2^+$ (5360 MeV, 80 MeV) for the AP1 potential.
They reside rather high up, excited by more than 500 MeV, from the lowest thresholds, $\Omega_c + D_s$ or $\Omega + J/\psi$.
Nevertheless, they all happen to be compact five-quark states, as is shown in the calculated density distribution of Fig.~\ref{r2rho}, whose coupling to baryon-meson scattering channels are weak.

Thus we conclude that the potential quark model predicts compact pentaquark resonances through the full five-body calculation. It would be interesting to observe such sharp and compact pentaquarks in future experiments. Simultaneous production of two charm and three strange quarks is generally very unlikely, but one may utilize bottom quark decay processes such as $\Xi_b^0 \rightarrow (sssc\bar{c})+K^+$ followed by the $(sssc\bar{c}) \rightarrow \Omega + J/\psi$ decay. Alternatively, high-energy heavy ion collisions are known to produce many strange and charm quarks, which can lead to the formation of pentaquark states. Resonance states may be observed in the $\Omega - J/\psi$ (or some other) correlations in the final states. If these states are observed, it would be a strong indication of the quark dynamics described by the quark model Hamiltonian, such as the quark confinement mechanism and spin dependent structures.

Another way of confirming our prediction is to investigate the resonance spectrum from lattice QCD calculations. As these pentaquarks contain only charm and strange quarks, we expect the reliability of the lattice simulation to be better than for systems with light ($u$ and $d$) quarks. 
Such a calculation is now in progress in our group.

\section*{Acknowledgments}

Q.M. is supported in part by the National Natural Science Foundation of China (under Grants No.11535005, No.\\11775118 and No.11690030) and the International Science $\&$ Technology Cooperation Program of China (under Grant No.2016YFE0129300). 
P.G. is supported by Grant-in-Aid for Early-Carrier Scientists No. JP18K13542 and the Leading Initiative for Excellent Young Researchers (LEADER) of the Japan Society for the Promotion of Science (JSPS). 
K.U.C is supported by the Special Postdoctoral Researcher (SPDR) program of RIKEN.
This work is supported in part by Grants-in-Aid for Scientific Research on Innovative Areas (No. JP18H05407, No. JP18H05236 (E.H and A.H.), No. JP19H05159 (M.O.)), and for Scientific Research No. JP17K05441(C) (A.H.). 

\printcredits

\bibliographystyle{elsarticle-num}

\bibliography{cas-refs}

\begin{thebibliography}{10}
\expandafter\ifx\csname url\endcsname\relax
  \def\url#1{\texttt{#1}}\fi
\expandafter\ifx\csname urlprefix\endcsname\relax\def\urlprefix{URL }\fi
\expandafter\ifx\csname href\endcsname\relax
  \def\href#1#2{#2} \def\path#1{#1}\fi

\bibitem{x3872}
S.-K. Choi, S.~Olsen, K.~Abe, T.~Abe, I.~Adachi, B.~S. Ahn, H.~Aihara, K.~Akai,
  M.~Akatsu, M.~Akemoto, et~al., Observation of a narrow charmoniumlike state
  in exclusive {$B^{\pm}\rightarrow K^{\pm} {\pi}^{+} {\pi}^{-} J/\psi$}
  decays, Physical Review Letters 91 (2003) 262001.
\newblock \href {http://dx.doi.org/10.1103/physrevlett.91.262001}
  {\path{doi:10.1103/physrevlett.91.262001}}.

\bibitem{Hosaka:2016pey}
A.~Hosaka, T.~Iijima, K.~Miyabayashi, Y.~Sakai, S.~Yasui, {Exotic hadrons with
  heavy flavors: X, Y, Z, and related states}, PTEP 2016~(6) (2016) 062C01.
\newblock \href {http://dx.doi.org/10.1093/ptep/ptw045}
  {\path{doi:10.1093/ptep/ptw045}}.

\bibitem{Aaij:2015tga}
R.~Aaij, et~al., {Observation of $J/\psi p$ Resonances Consistent with
  Pentaquark States in $\Lambda_b^0 \to J/\psi K^- p$ Decays}, Phys. Rev. Lett.
  115 (2015) 072001.
\newblock \href {http://dx.doi.org/10.1103/PhysRevLett.115.072001}
  {\path{doi:10.1103/PhysRevLett.115.072001}}.

\bibitem{pc2019}
R.~Aaij, C.~A. Beteta, B.~Adeva, M.~Adinolfi, C.~A. Aidala, Z.~Ajaltouni,
  S.~Akar, P.~Albicocco, J.~Albrecht, F.~Alessio, et~al., Observation of a
  narrow pentaquark state, ${P_c}(4312)^+$, and of the two-peak structure of
  the ${P_c}(4450)^+$, Physical Review Letters 122~(22) (2019) 222001.
\newblock \href {http://dx.doi.org/10.1103/PhysRevLett.122.222001}
  {\path{doi:10.1103/PhysRevLett.122.222001}}.

\bibitem{Maiani:2015vwa}
L.~Maiani, A.~D. Polosa, V.~Riquer, {The New Pentaquarks in the Diquark Model},
  Phys. Lett. B749 (2015) 289--291.
\newblock \href {http://dx.doi.org/10.1016/j.physletb.2015.08.008}
  {\path{doi:10.1016/j.physletb.2015.08.008}}.

\bibitem{Lebed:2015tna}
R.~F. Lebed, {The Pentaquark Candidates in the Dynamical Diquark Picture},
  Phys. Lett. B749 (2015) 454--457.
\newblock \href {http://dx.doi.org/10.1016/j.physletb.2015.08.032}
  {\path{doi:10.1016/j.physletb.2015.08.032}}.

\bibitem{Wang:2015epa}
Z.-G. Wang, {Analysis of $P_c(4380)$ and $P_c(4450)$ as pentaquark states in
  the diquark model with QCD sum rules}, Eur. Phys. J. C76~(2) (2016) 70.
\newblock \href {http://dx.doi.org/10.1140/epjc/s10052-016-3920-4}
  {\path{doi:10.1140/epjc/s10052-016-3920-4}}.

\bibitem{Li:2015gta}
G.-N. Li, X.-G. He, M.~He, {Some Predictions of Diquark Model for Hidden Charm
  Pentaquark Discovered at the LHCb}, JHEP 12 (2015) 128.
\newblock \href {http://dx.doi.org/10.1007/JHEP12(2015)128}
  {\path{doi:10.1007/JHEP12(2015)128}}.

\bibitem{Takeuchi:2016ejt}
S.~Takeuchi, M.~Takizawa, {The hidden charm pentaquarks are the hidden
  color-octet $uud$ baryons?}, Phys. Lett. B764 (2017) 254--259.
\newblock \href {http://dx.doi.org/10.1016/j.physletb.2016.11.034}
  {\path{doi:10.1016/j.physletb.2016.11.034}}.

\bibitem{Roca:2015dva}
L.~Roca, J.~Nieves, E.~Oset, {LHCb pentaquark as a
  $\bar{D}^*\Sigma_c-\bar{D}^*\Sigma_c^*$ molecular state}, Phys. Rev. D92~(9)
  (2015) 094003.
\newblock \href {http://dx.doi.org/10.1103/PhysRevD.92.094003}
  {\path{doi:10.1103/PhysRevD.92.094003}}.

\bibitem{He:2015cea}
J.~He, {$\bar{D}\Sigma^*_c$ and $\bar{D}^*\Sigma_c$ interactions and the LHCb
  hidden-charmed pentaquarks}, Phys. Lett. B753 (2016) 547--551.
\newblock \href {http://dx.doi.org/10.1016/j.physletb.2015.12.071}
  {\path{doi:10.1016/j.physletb.2015.12.071}}.

\bibitem{Xiao:2015fia}
C.~W. Xiao, U.~G. Meissner, {$J/\psi(\eta_c)N$ and $\Upsilon(\eta_b)N$ cross
  sections}, Phys. Rev. D92~(11) (2015) 114002.
\newblock \href {http://dx.doi.org/10.1103/PhysRevD.92.114002}
  {\path{doi:10.1103/PhysRevD.92.114002}}.

\bibitem{Chen:2015loa}
R.~Chen, X.~Liu, X.-Q. Li, S.-L. Zhu, {Identifying exotic hidden-charm
  pentaquarks}, Phys. Rev. Lett. 115~(13) (2015) 132002.
\newblock \href {http://dx.doi.org/10.1103/PhysRevLett.115.132002}
  {\path{doi:10.1103/PhysRevLett.115.132002}}.

\bibitem{Liu:2019tjn}
M.-Z. Liu, Y.-W. Pan, F.-Z. Peng,
  M.~S$\mathrm{\acute{a}}$nchez~S$\mathrm{\acute{a}}$nchez, L.-S. Geng,
  A.~Hosaka, M.~Pavon~Valderrama, {Emergence of a complete heavy-quark spin
  symmetry multiplet: seven molecular pentaquarks in light of the latest LHCb
  analysis}, Phys. Rev. Lett. 122 (2019) 242001.
\newblock \href {http://dx.doi.org/10.1103/PhysRevLett.122.242001}
  {\path{doi:10.1103/PhysRevLett.122.242001}}.

\bibitem{Yamaguchi:2017zmn}
Y.~Yamaguchi, A.~Giachino, A.~Hosaka, E.~Santopinto, S.~Takeuchi, M.~Takizawa,
  {Hidden-charm and bottom meson-baryon molecules coupled with five-quark
  states}, Phys. Rev. D96~(11) (2017) 114031.
\newblock \href {http://dx.doi.org/10.1103/PhysRevD.96.114031}
  {\path{doi:10.1103/PhysRevD.96.114031}}.

\bibitem{Kubarovsky:2015aaa}
V.~Kubarovsky, M.~B. Voloshin, {Formation of hidden-charm pentaquarks in
  photon-nucleon collisions}, Phys. Rev. D92~(3) (2015) 031502.
\newblock \href {http://dx.doi.org/10.1103/PhysRevD.92.031502}
  {\path{doi:10.1103/PhysRevD.92.031502}}.

\bibitem{Esposito:2015fsa}
A.~Esposito, A.~L. Guerrieri, L.~Maiani, F.~Piccinini, A.~Pilloni, A.~D.
  Polosa, V.~Riquer, {Observation of light nuclei at ALICE and the X(3872)
  conundrum}, Phys. Rev. D92~(3) (2015) 034028.
\newblock \href {http://dx.doi.org/10.1103/PhysRevD.92.034028}
  {\path{doi:10.1103/PhysRevD.92.034028}}.

\bibitem{hiyama2006five}
E.~Hiyama, M.~Kamimura, A.~Hosaka, H.~Toki, M.~Yahiro, Five-body calculation of
  resonance and scattering states of pentaquark system, Physics Letters B 633
  (2006) 237--244.
\newblock \href {http://dx.doi.org/10.1016/j.physletb.2005.11.086}
  {\path{doi:10.1016/j.physletb.2005.11.086}}.

\bibitem{hiyama2018five}
E.~Hiyama, A.~Hosaka, M.~Oka, J.~M. Richard, Quark model estimate of
  hidden-charm pentaquark resonances, Physical Review C 98~(4) (2018) 1--8.
\newblock \href {http://dx.doi.org/10.1103/PhysRevC.98.045208}
  {\path{doi:10.1103/PhysRevC.98.045208}}.

\bibitem{Padmanath:2019wid}
M.~Padmanath, {Hadron Spectroscopy and Resonances: Review}, PoS LATTICE2018
  (2018) 013.
\newblock \href {http://dx.doi.org/10.22323/1.334.0013}
  {\path{doi:10.22323/1.334.0013}}.

\bibitem{silvestre1996spectrum}
B.~Silvestre-Brac, Spectrum and static properties of heavy baryons, Few-Body
  Systems 20 (1996) 1--25.
\newblock \href {http://dx.doi.org/10.1007/s006010050028}
  {\path{doi:10.1007/s006010050028}}.

\bibitem{semay1994diquonia}
C.~Semay, B.~Silvestre-Brac, Diquonia and potential models, Zeitschrift f{\"u}r
  Physik C Particles and Fields 61 (1994) 271--275.
\newblock \href {http://dx.doi.org/10.1007/BF01413104}
  {\path{doi:10.1007/BF01413104}}.

\bibitem{GEM1988}
M.~Kamimura, Nonadiabatic coupled-rearrangement-channel approach to muonic
  molecules, Physical Review A 38 (1988) 621.
\newblock \href {http://dx.doi.org/10.1103/PhysRevA.38.621}
  {\path{doi:10.1103/PhysRevA.38.621}}.

\bibitem{GEM2003}
E.~Hiyama, Y.~Kino, M.~Kamimura, Gaussian expansion method for few-body
  systems, Progress in Particle and Nuclear Physics 51 (2003) 223--307.
\newblock \href {http://dx.doi.org/10.1016/S0146-6410(03)90015-9}
  {\path{doi:10.1016/S0146-6410(03)90015-9}}.

\bibitem{hiyama2000lambda}
E.~Hiyama, M.~Kamimura, T.~Motoba, T.~Yamada, Y.~Yamamoto, {$\Lambda$ N}
  spin-orbit splittings in $^9_{\Lambda}{Be}$ and $^{13}_{\Lambda}{Be}$ studied
  with one-boson-exchange {$\Lambda$ N} interactions, Physical review letters
  85 (2000) 270.
\newblock \href {http://dx.doi.org/10.1103/PhysRevLett.85.270}
  {\path{doi:10.1103/PhysRevLett.85.270}}.

\bibitem{hiyama2004four}
E.~Hiyama, B.~Gibson, M.~Kamimura, Four-body calculation of the first excited
  state of $^4{He}$ using a realistic $nn$ interaction: $^4{He}(e,e')$
  $^4{He}(0^+_2)$ and the monopole sum rule, Physical Review C 70 (2004)
  031001.
\newblock \href {http://dx.doi.org/10.1103/PhysRevC.70.031001}
  {\path{doi:10.1103/PhysRevC.70.031001}}.

\bibitem{hiyama2009structure}
E.~Hiyama, T.~Yamada, Structure of light hypernuclei, Progress in Particle and
  Nuclear Physics 63 (2009) 339--395.
\newblock \href {http://dx.doi.org/10.1016/j.ppnp.2009.05.001}
  {\path{doi:10.1016/j.ppnp.2009.05.001}}.

\bibitem{hiyama2015resonant}
E.~Hiyama, M.~Isaka, M.~Kamimura, T.~Myo, T.~Motoba, Resonant states of the
  neutron-rich {$\Lambda$} hypernucleus $^7_{\Lambda}{Be}$, Physical Review C
  91 (2015) 054316.
\newblock \href {http://dx.doi.org/10.1103/PhysRevC.91.054316}
  {\path{doi:10.1103/PhysRevC.91.054316}}.

\bibitem{brink1998tetraquarks}
D.~Brink, F.~Stancu, Tetraquarks with heavy flavors, Physical Review D 57
  (1998) 6778.
\newblock \href {http://dx.doi.org/10.1103/PhysRevD.57.6778}
  {\path{doi:10.1103/PhysRevD.57.6778}}.

\bibitem{1981stablization}
J.~Simons, Resonance state lifetimes from stabilization graphs, The Journal of
  Chemical Physics 75 (1981) 2465--2467.
\newblock \href {http://dx.doi.org/10.1063/1.442271}
  {\path{doi:10.1063/1.442271}}.

\end{thebibliography}

\end{document}